\documentclass[letterpaper]{article} 
\usepackage{aaai25}  
\usepackage{times}  
\usepackage{helvet}  
\usepackage{courier}  
\usepackage[hyphens]{url}  
\usepackage{graphicx} 
\usepackage{natbib}  
\usepackage{caption} 
\usepackage{algorithm}
\usepackage{algorithmic}

\usepackage{booktabs}
\usepackage{subcaption}
\usepackage{multirow}

\newcommand{\x}{\bm{x}}

\newcommand{\h}{\bm{h}}

\newcommand{\I}{\mathbf{I}}
\newcommand{\C}{\mathbf{C}}

\newcommand{\W}{\mathbf{W}}

\newcommand{\F}{\mathbf{F}}

\newcommand{\G}{\mathbf{G}}

\usepackage{amsmath,amsfonts,bm}

\def\eqref#1{equation~\ref{#1}}

\def\1{\bm{1}}

\DeclareMathAlphabet{\mathsfit}{\encodingdefault}{\sfdefault}{m}{sl}
\SetMathAlphabet{\mathsfit}{bold}{\encodingdefault}{\sfdefault}{bx}{n}

\usepackage{newfloat}
\usepackage{listings}
\DeclareCaptionStyle{ruled}{labelfont=normalfont,labelsep=colon,strut=off} 
\floatstyle{ruled}
\newfloat{listing}{tb}{lst}{}
\floatname{listing}{Listing}
\title{DP-MemArc: Differential Privacy Transfer Learning for Memory Efficient Language Models}

\author{
    Yanming Liu\textsuperscript{\rm 1},
    Xinyue Peng\textsuperscript{\rm 2},
    Yuwei Zhang\textsuperscript{\rm 8},
    Xiaolan Ke\textsuperscript{\rm 3},
    Songhang Deng\textsuperscript{\rm 4},
    Jiannan Cao\textsuperscript{\rm 5}, 
    Chen Ma\textsuperscript{\rm 6},
    Mengchen Fu\textsuperscript{\rm 7},
    Tianyu Du\textsuperscript{\rm 1}\thanks{Corresponding Author.},
    Sheng Cheng\textsuperscript{\rm 1},
    Xun Wang\textsuperscript{\rm 1},
    Jianwei Yin\textsuperscript{\rm 1},
    Xuhong Zhang\textsuperscript{\rm 1}\textsuperscript{*} 
}

\affiliations {
    \textsuperscript{\rm 1}Zhejiang University\\ 
    \textsuperscript{\rm 2}Southeast University\\
    \textsuperscript{\rm 3}Harvard University\\
    \textsuperscript{\rm 4}University of California, Los Angeles\\
    \textsuperscript{\rm 5}Massachusetts Institute of Technology\\
    \textsuperscript{\rm 6}Renmin University of China\\
    \textsuperscript{\rm 7}The University of Tokyo\\
    \textsuperscript{\rm 8}Tongji University\\
    \{oceann24, zhangxuhong, zjradty, zjuyjw\}@zju.edu.cn, 
    xinyuepeng@seu.edu.cn, 
    jiannan@mit.edu, songh00@ucla.edu
}

\begin{document}

\maketitle

\begin{abstract}
Large language models have repeatedly shown outstanding performance across diverse applications. However, deploying these models can inadvertently risk user privacy. The significant memory demands during training pose a major challenge in terms of resource consumption. This substantial size places a heavy load on memory resources, raising considerable practical concerns. In this paper, we introduce DP-MemArc, a novel training framework aimed at reducing the memory costs of large language models while emphasizing the protection of user data privacy. DP-MemArc incorporates side network or reversible network designs to support a variety of differential privacy memory-efficient fine-tuning schemes. Our approach not only achieves about 2.5 times in memory optimization but also ensures robust privacy protection, keeping user data secure and confidential. Extensive experiments have demonstrated that DP-MemArc effectively provides differential privacy-efficient fine-tuning across different task scenarios.
\end{abstract}

%

\section{Introduction}

Large language models (LLMs)~\cite{radford2019language, hoffmann2022training, chowdhery2023palm, touvron2023llama} have already demonstrated their capabilities across various domains, excelling in a wide range of generation and comprehension tasks~\cite{bang2023multitask,robinson2022leveraging,li2022self}. However, complete training of LLMs demands significant computational resources and time, making it inconvenient to adapt the model in downstream tasks~\cite{liu2022few}. There exist several methods that offer solutions for parameter-efficient fine-tuning \cite{dettmers2024qlora, houlsby2019parameter, hu2021lora}. These approaches achieve highly effective downstream task fine-tuning results by adjusting only a small number of parameters. The goal of such methods is to enable LLMs to adapt to small-scale features in a relatively small dataset, thereby accomplishing specific downstream tasks. Unfortunately, for LLMs, we often encounter situations where the available dataset is small and proprietary, raising concerns about privacy \cite{bu2024differentially, yu2021differentially, finlayson2024logits}. Additionally, the training of LLMs requires substantial training memory \cite{wang2023read}, making it challenging to train on parameter-efficient fine-tuning.

\begin{table*}[]
    \centering
    \setlength\tabcolsep{2pt}
    \resizebox{\linewidth}{!}{    
    \begin{tabular}{|c|c|c|c|c|c|}
    \toprule
     \multirow{2}{*}{Module} &\multirow{2}{*}{Forward pass} &\multirow{2}{*}{Back-propagation}&\multirow{2}{*}{\shortstack[c]{Ghost norm in \\ Book-Keeping}} &\multirow{2}{*}{\shortstack[c]{Opacus grad\\ instantiation}}&\multirow{2}{*}{\shortstack[c]{Opacus sum of \\weighted grad}} \\
        &&&&& \\ \midrule
        Time complexity &$2BTpd$&$4BTpd$&$2BT^2(p+d)$&$2BTpd$&$2Bpd$ \\ \midrule
        Space complexity&$pd+BTd$&$BT(p+d)+pd$&$2BT^2$&$Bpd$&0 \\ \bottomrule
    \end{tabular}
    }
    \caption{The time and space complexity of the training process of a model under a single-layer MLP. While opacus is a codebase for vanilla differential private method implementation.}
    \label{tab:block complexity}
\end{table*}

A recent line of work that focuses on fine-tuning large models using differential privacy (DP) solutions, including both full parameter fine-tuning and parameter efficient fine-tuning approaches \cite{duan2024flocks, bu2024differentially, yu2021differentially}. These solutions employ a method called Differential Privacy Stochastic Gradient Descent (DP-SGD) \cite{yu2019differentially}. The training data is protected by implementing gradient clipping and adding Gaussian noise during each iteration to ensure privacy. Compared to traditional fine-tuning approaches, DP allows for downstream task handling with only a small loss in accuracy while maintaining a theoretical private guarantee \cite{yu2021differentially}. These approaches exhibit good performance across a variety of tasks and settings. However, these methods still have issues with training memory. In previous research, differential privacy has imposed larger computational and storage overheads, making training such large models challenging in resource-constrained scenarios. Additionally, existing efficient parameter fine-tuning with differential privacy schemes has only achieved marginal reductions in memory overhead, with insufficient optimization efficiency in memory resources \cite{li2022does, ke2024convergence}. As models continue to grow, the demand for both memory efficiency and privacy in such scenarios also increases.

To address this issue, we propose a solution called \textbf{D}ifferential \textbf{P}rivate \textbf{Mem}ory efficient transfer \textbf{Arc}hitecture (\textbf{DP-MemArc}), a framework for training in scenarios that involve both privacy-protection and memory efficient transfer learning. In our framework, we explore two efficient methods for parameter-efficient fine-tuning, DP-MemArc$_\text{side}$ and DP-MemArc$_\text{rev}$, which save memory usage from different perspectives. In this setup, our approach not only achieves competitive performance but also significantly reduces training memory usage. Experiments on different datasets and models have thoroughly demonstrated the effectiveness and potential of our approach. Therefore, our work effectively addresses the issue of insufficient memory in private fine-tuning for language models, while also providing alternative privacy fine-tuning solutions.

In summary, our contributions in this paper are as follows:

\begin{itemize}
\item We propose a framework called DP-MemArc, which enables efficient fine-tuning of language models with lower training memory in differential privacy fine-tuning. This framework contains two memory optimization methods, reducing the memory requirements for privacy training of language models.
\end{itemize}

\begin{itemize}
\item We conduct a systematic analysis of the relationship between training memory requirements and network architecture. We elucidate the characteristics of fine-tuning memory cost under different network architectures, demonstrating favorable downstream task performance in differential privacy.
\end{itemize}

\begin{itemize}
\item We evaluate our DP-MemArc framework on multiple datasets and models. The results show promising performance across various dimensions, with a substantial improvement in training memory.
\end{itemize}

\section{Preliminaries}

\subsection{Memory Footprint on Language Model}

We consider a $N$ multilayer perception: $\boldsymbol{x}_N = f_N(f_{N-1}(...f_2(f_1(\boldsymbol{x}_0))))$, where $x_0$ as the initial PLM input, the $i^{th}$ layer $\boldsymbol{x}_i = f_i(\boldsymbol{x}_{i-1}) = \sigma_i(\boldsymbol{\W}_i\boldsymbol{x}_{i-1}$) consists of a weight matrix $\W_i$ and a nonelinear function $\boldsymbol{\sigma}_i$. For the format simplicity, the bias term is ignored. We denote $\boldsymbol{h}_i = \boldsymbol{\W}_i\boldsymbol{x}_{i-1}$ as the hidden states for the pre-activation of $i^{th}$ layer. In backpropagation with loss $\mathcal{L}$, the gradient of $W_i$ is calculated with  respect to $x_i$ using the chain rule:

\setlength{\abovedisplayskip}{0pt}

\begin{equation}
     \frac{\partial{\mathcal{L}}}{\partial{\boldsymbol{\W}_i}} = \frac{\partial{\mathcal{L}}}{\partial{\boldsymbol{x}_N}}(\prod \limits_{j=i+1}^N\frac{\partial{\boldsymbol{x}_j}}{\partial{\boldsymbol{h}_j}}\frac{\partial{\boldsymbol{h}_j}}{\partial{\boldsymbol{x}_{j-1}}})\frac{\partial{\boldsymbol{x}_i}}{\partial{\boldsymbol{h}_i}}\frac{\partial{\boldsymbol{h}_i}}{\partial{\boldsymbol{\W}_i}}
\end{equation}

Denoting the derivative of $\boldsymbol{\sigma}$ is $\boldsymbol{\sigma}'$, then the equation could simplified as:

\begin{equation}
\frac{\partial{\mathcal{L}}}{\partial{\boldsymbol{\W}_i}} = 
 \frac{\partial{\mathcal{L}}}{\partial{\boldsymbol{x}_N}}(\prod \limits_{j=i+1}^N\boldsymbol{\sigma}_j'\boldsymbol{\W}_j)\boldsymbol{\sigma}'_{i}\boldsymbol{x}_{i-1}.
\end{equation}

Thus, in training memory, the core consumption lies in the states of model weights $\{\W_i\}_{i=1}^N$ and derivative activation functions state $\{\boldsymbol{\sigma}'\}_{i=1}^N$ along the backpropagation path, as well as the optimizer states used during gradient updates. The optimizer states are directly related to the updated model parameters $\{\Delta \W\}$.

Assuming the batch size is $B$, the length of the input sequence is $T$, the model input and output dimension is $d$ and $p$, for a standard linear layer $\boldsymbol{x}_i = \sigma_i(\boldsymbol{\W}_i\boldsymbol{x}_{i-1})$, the forward pass stores the intermediate states of the model and the model weights with the memory complexity of $O(pd + BTd)$, while the backward pass is responsible for storing the activation states during the gradient update process, the results of the output gradients, and the corresponding parameter gradients, with the total memory complexity of $O(BT(p+d) + pd)$.

\subsection{Deep Learning with Differential Privacy} 

Differential Privacy \cite{dwork2006calibrating, abadi2016deep} algorithms demonstrate that under this formulation, the model's output cannot significantly help determine whether an individual record exists in the input dataset through certain mathematical derivations. The formal definition is recalled as follows:

\newtheorem{definition}{Definition}
\begin{definition}
(Differential Privacy). Given a domain $\mathcal{D}$, any two datasets $D$, $D' \subseteq \mathcal{D}$ that differ in exactly one record are called neighboring datasets. A randomized algorithm $\mathcal{M} : \mathcal{D} \rightarrow \mathcal{R}$ is $(\epsilon,\delta)$-differential private if for all neighboring datasets $D$ and $D'$ and all $T \subseteq \mathcal{R}$,

\vspace{0pt}
\setlength{\abovedisplayskip}{0pt}

$$\Pr [\mathcal{M}(D) \subseteq T] \leq e^{\epsilon} \Pr [\mathcal{M}(\mathcal{D}') \subseteq T] + \delta. $$

\end{definition}

{
\setlength{\parindent}{0cm}

\textbf{DP-optimizer.} To train a privacy-preserving language model, the current approach involves providing differential privacy guarantees when computing gradients and applying these guarantees to optimizers such as SGD or Adam \cite{abadi2016deep, mironov2017renyi, koskela2020computing}. This approach incorporates steps involving per-example gradient clipping $\G_l=\sum C_i\frac{\partial \mathcal{L}_i}{\partial \W_{(l)}}$ and adding Gaussian noise $\mathcal{N}(0, \I)$ on gradient $\G$. Where $C_i$ is the per-sample clipping factor.

\vspace{0.2cm}

\textbf{Book-keeping.} To avoid the significant memory overhead caused by storing gradients for each sample during initialization, \citet{bu2023differentially} proposed a method BK utilizing gradient norms. Using the GhostClip \cite{goodfellow2015efficient, bu2022scalable} strategy, the gradient norm of each sample is calculated. 

\setlength{\abovedisplayskip}{0pt}

\begin{align}
\begin{split}
&\left\|\frac{\partial \mathcal{L}_i}{\partial \W}\right\|_\text{F}^2=
\text{vec}\Big(\frac{\partial \mathcal{L}}{\partial \h_i} \frac{\partial \mathcal{L}}{\partial \h_i}^\top\Big)\cdot\text{vec}\left(\x_i \x_i^\top\right)
\end{split}
\label{eq:ghost norm}
\end{align}

}

Based on the gradient norms, clipping factors $C_i$ and clipping matrices $\C$ are generated, which are then used to compute the sum of clipped gradient in batches $\G_l=\x_{(l)}^\top\text{diag}(\C)\frac{\partial \mathcal{L}}{\partial \h_{(l)}}$. It is necessary to retain the complexity of the original DP to avoid issues related to length-dependent time and memory complexity when handling long-text question answering. BK-MixOpt strikes a balance between the two. It compares the theoretical complexity in terms of both dimension and context length, and selects the optimal memory complexity $O(min\{2BT^2, Bpd\})$ as the basis for computation.

\begin{figure*}[htbp]
    \centering
    \begin{subfigure}{0.72\textwidth}
        \centering
        \includegraphics[width=\textwidth]{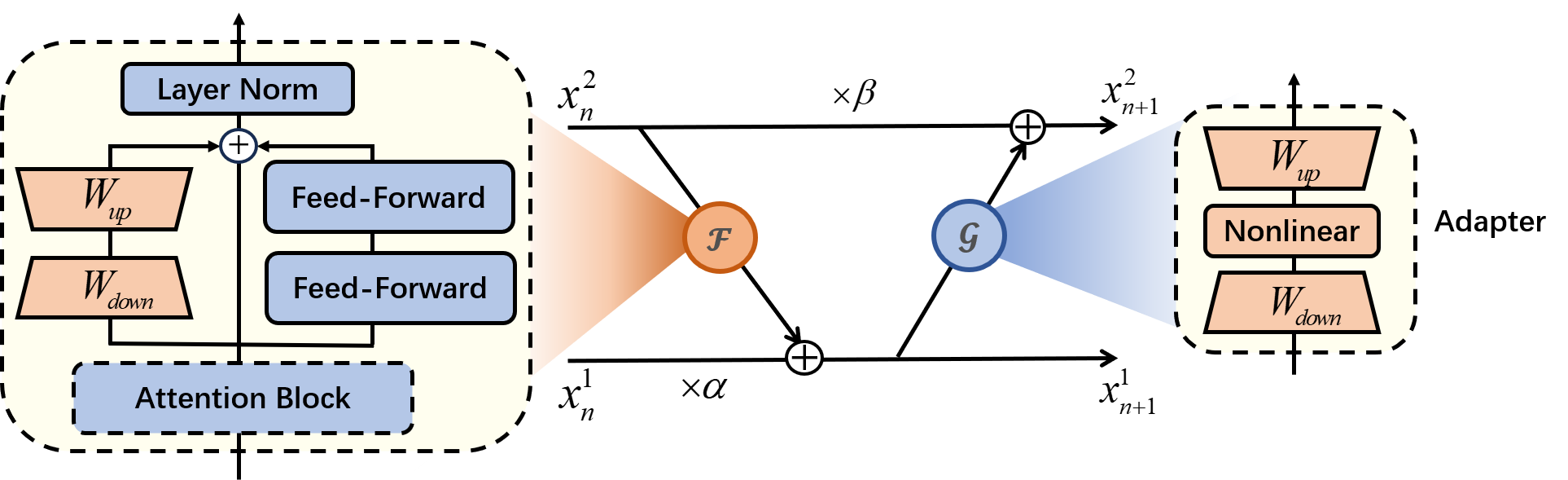}
        \caption{Designs of DP-MemArc$_\text{rev}$.}
        \label{fig:image1}
    \end{subfigure}
    \hfill
    \begin{subfigure}{0.26\textwidth}
        \centering
        \includegraphics[width=\textwidth]{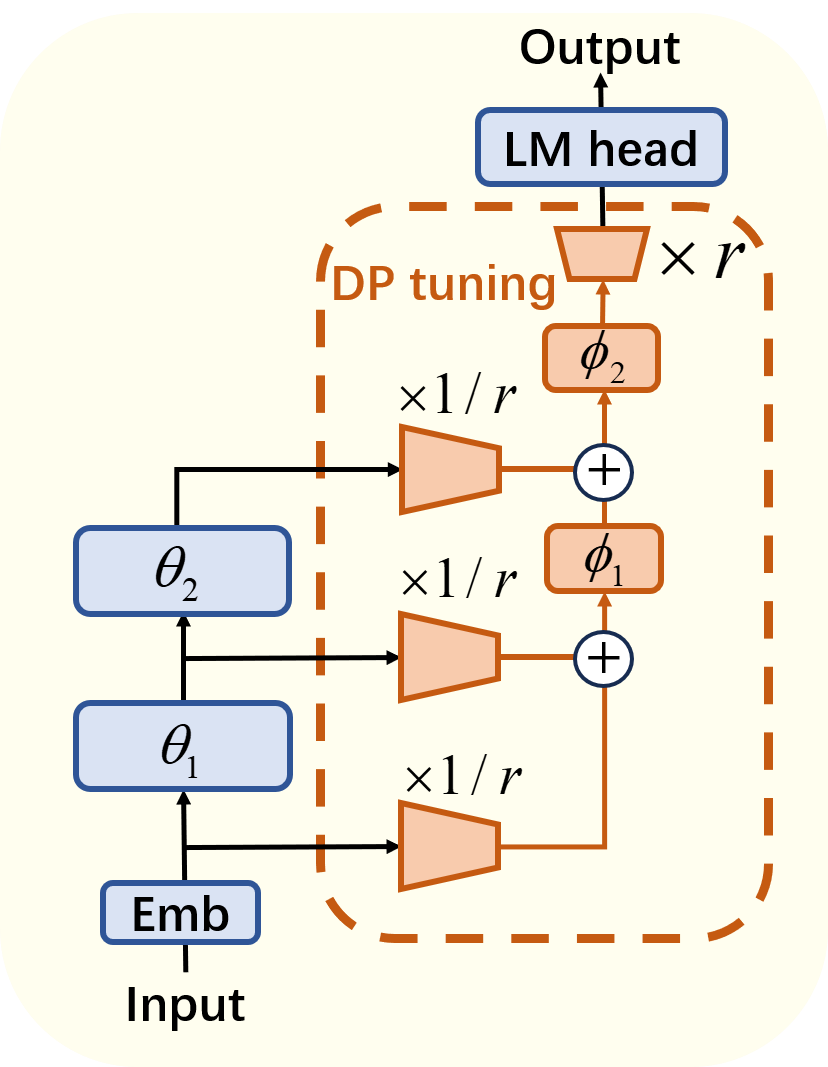}
        \caption{Designs of DP-MemArc$_\text{side}$.}
        \label{fig:image2}
    \end{subfigure}
    \caption{Two different DP-MemArc designs, the left represents reversible network design, and the right represents side network design. The trainable parameters are fine-tuned using the differential privacy BK-MixOpt method.}
    \label{fig:combined}
\end{figure*}

\section{Methodology}

To address the issue of excessive memory consumption during differential privacy training, we have designed two methods: $\text{DP-MemArc}_\text{side}$ and $\text{DP-MemArc}_\text{rev}$. These methods help reduce training memory usage in different aspects.

\subsection{Side Network Design}

In general, most of the memory consumption comes from the model weights and the states of activation functions in the backward propagation path. By minimizing the consumption of these two parts as much as possible, the memory usage during training can be correspondingly reduced. This necessitates finding a reasonable design to address this situation.

Assume the base model is $\F$, the model's pre-trained weights, input, output, and parameters are $\W_\text{p}$, $x_0$, $y$, $\theta$. The model could formulated as:

\setlength{\abovedisplayskip}{0pt}

\begin{align}
\begin{split}
& y = \F(\W_\text{p}, \theta; x_0).
\end{split}
\label{eq:side_func1}
\end{align}

Traditional parameter-efficient fine-tuning methods cannot avoid the memory consumption associated with the model weights of frozen parameters in the backward propagation path, which can formulated as:

\setlength{\abovedisplayskip}{0pt}

\begin{align}
\begin{split}
& y = \F(\W_\text{p} + \Delta\W, \theta + \Delta\theta; x_0).
\end{split}
\label{eq:side_func2}
\end{align}

We hope to find a form that remains distinct from the original form when adjusting parameters. That is, there exists such a form:

\setlength{\abovedisplayskip}{0pt}

\begin{align}
\begin{split}
y = \alpha\F_1(\W_\text{p} , \theta; x_0) + \beta \F_2(\Delta\W , \Delta\theta; x_0).
\end{split}
\label{eq:side_func3}
\end{align}

In this form, Side-tuning \cite{zhang2020side} meets the requirements. Side-tuning introduces a side network that learns the knowledge and features of new tasks, relying on the knowledge contained in the trained model parameters, thus supporting the processing of downstream tasks.

Assuming the input and output dimension of the side network is $r$, we add a liner layer at the last layer of the side network to ensure dimension consistency. We use Book-keeping \cite{bu2023differentially} for differential privacy fine-tuning. The memory cost includes both the forward and backward propagation processes. 
For the forward process, the bilateral forward propagation memory consumption needs to be taken into account, with the complexity of $O(pd+r^2+BT(d+r))$. For the backward process, gradients need to be computed only in the side network, with a complexity of $O(2BTr + r^2)$ for gradients and $O(2BT^2)$ for Ghost Norm. 

When $r \ll d$, the side tuning approach significantly reduces the training memory required for privacy fine-tuning. However, at sufficiently small $r$, the performance of side tuning also deteriorates significantly. To integrate the information of a trained model effectively into side networks, we adopt the LST \cite{sung2022lst} method. This involves passing the intermediate layer information from the pre-trained model to the si de through a linear layer $f'$. We denote this method as DP-MemArc$_\text{side}$.

\setlength{\abovedisplayskip}{0pt}

\begin{align}
\begin{split}
y =  \F_2(\Delta\W , \Delta\theta; y_{i-1} + f_{i}'(x_i), y_0=x_0).
\end{split}
\label{eq:side_func}
\end{align}


Using the DP-MemArc$_\text{side}$ method, we can maintain a good performance in fine-tuning our side network with differential privacy. When $d / r= 8$, LST \cite{sung2022lst} and DP-MemArc$_\text{side}$ achieves an empirically optimal ratio of training memory to performance.

\subsection{Reversible Network Design}
\label{sec:reverse}

Due to the significant amount of training memory required to store the state of activation functions during batch processing, a large portion of memory is consumed by saving activation states $\{\boldsymbol{\sigma_i}\}_{i=1}^N$. Regular parameter-efficient fine-tuning methods cannot effectively address this issue. DP-MemArc$_\text{side}$ reduces the memory needed to store activation functions by compressing the dimensions of the activation functions. However, this method still consumes some memory. If we could deduce the intermediate states from the output results in reverse, we could further reduce the memory demand for storing activation states.

For reversible networks \cite{gomez2017reversible, liao2023make}, the following form is usually satisfied.

\begin{flalign}
 \begin{aligned}[c]
    &\bm{x}_{i+1}^1 = \alpha \bm{x}_i^1 + \mathcal{F}_i(\bm{x}_i^2), \\
    &\bm{x}_{i+1}^2 = \beta \bm{x}_i^2 + \mathcal{G}_i(\bm{x}_{i+1}^1), \\
    &\bm{x}_i^2 = (\bm{x}_{i+1}^2 - \mathcal{G}_i(\bm{x}_{i+1}^1)) / \beta, \\
    &\bm{x}_i^1 = (\bm{x}_{i+1}^1 - \mathcal{F}_i(\bm{x}_i^2)) / \alpha  .
 \end{aligned}
\end{flalign}

We can obtain the corresponding activation function values $\boldsymbol{\sigma_i} = \sigma_i(\boldsymbol{\W}_i\boldsymbol{x}_{i-1})$ from the intermediate states $\{\boldsymbol{x_i}\}_{i=1}^N$ of the model and calculate their derivatives, thus avoiding the need to store each activation function value.

To enable the two modules of the reversible network to both acquire new features and retain the knowledge of the pre-trained model, For module $\mathcal{F}$, we introduced the LoRA \cite{hu2021lora} architecture into the FFN layer of the model, continuing the traditional LoRA approach. Meanwhile, for module $\mathcal{G}$, we used Adapters \cite{houlsby2019parameter} as trainable parameters to adapt to downstream tasks. Since the network is reversible, we only need to use constant reproducible space to compute $\bm{x}_i^2$ and $\bm{x}_i^1$ for each layer, which satisfies the requirements for the subsequent backpropagation calculations. We denote this method as DP-MemArc$_\text{rev}$.

For reversible networks, we have the following derivation steps. 
At the beginning of training, when the output of the adapter output is close to 0. $x_n \approx \mathcal{F}_n(\x_{n-1})$. Assume that $x_0^1$ and $x_0^2$ comes from the initial input $x$, we have:

\begin{align}
 \begin{aligned}[c]
    \x_1^1 &= \alpha \x_0^1 + \mathcal{F}_1(\x_0^2) \approx \alpha x_0 + x_1, \\
 \end{aligned}
\end{align}

\begin{align}
 \begin{aligned}[c]
    \x_1^2 &= \beta \x_0^2 + \mathcal{G}_1(\x_1^1) = \beta x_0 + \mathcal{G}_1(\x_1^1)  \approx \beta \x_0. \\
 \end{aligned}
\end{align}

When $\alpha \rightarrow 0$, we have $\x_1^1 = x_1$, $\x_1^2 = \beta x_0$. We achieve a relatively stable state of the reversible network by exchanging output values $\x_1^1 = \beta \x_0$, $\x_1^2 = x_1$. Through iterative computation like this, the model can be satisfied as $\x_n^1 \approx \beta \x_{n-1}$, $\x_n^2 \approx x_n$. We generate the final output as $x = (\x_N^1 + \x_N^2)/2$. In this way, when training reversible models, the continuity of the model's representation can be maintained, and inference and learning for downstream tasks can be facilitated based on pre-trained models.

During the backpropagation process in our reversible network, the intermediate states of the model can be obtained by computing the reverse steps. As a result, the training memory required for activation values can be reduced by reusing a fixed-size replaceable memory. The primary training memory consumption of the model comes from storing the output gradients, storing the parameter gradients, and the computational memory required by the Ghostnorm method. Here, we also employ the BK-MixOpt algorithm to calculate the norm of the samples, thereby obtaining the corresponding gradient values. During training, we set batch sizes to 32.

\begin{table*}[!t]
    \resizebox{\linewidth}{!}{    
\begin{tabular}{@{}c|ccccccccccccccc|c@{}}
\toprule
\multirow{2}{*}{}                                            & \multicolumn{3}{c}{Memory(GB)$\downarrow$}                                                              & \multicolumn{3}{c}{MNLI$\uparrow$}                                                                                                               & \multicolumn{3}{c}{QQP$\uparrow$}                                                                                                                & \multicolumn{3}{c}{QNLI$\uparrow$}                                                                                                               & \multicolumn{3}{c}{SST2$\uparrow$}    & \multicolumn{1}{|c}{\multirow{2}{*}{Trainable param(\%)}}                                                                \\ \cmidrule(lr){2-4} \cmidrule(lr){5-7} \cmidrule(lr){8-10} \cmidrule(lr){11-13} \cmidrule(lr){14-16} 
                                                             & $\epsilon = 1.6$ & $\epsilon = 8$ & $\epsilon = \infty$ & $\epsilon = 1.6$ & $\epsilon = 8$ & $\epsilon = \infty$ & $\epsilon = 1.6$ & $\epsilon = 8$ & $\epsilon = \infty$ & $\epsilon = 1.6$ & $\epsilon = 8$ & $\epsilon = \infty$ & $\epsilon = 1.6$ & $\epsilon = 8$ & $\epsilon = \infty$ & \\ \midrule
\multicolumn{17}{c}{\textbf{\textit{$\text{Differential Private on Adaptive Parameter Transfer Learning}$}}} \\ \midrule

DP-LoRA                                                      &       12.65                      &     12.21                        &      7.14                           &        \textbf{81.49}                     &       87.07                      & 90.81                                                                           &       83.46                      &      88.53                       &                                     \textbf{91.75}                                       &             \textbf{87.32}                &         91.34                    &         94.33                                                                   &      93.43                       &      95.14                       &   95.88                              &1.88\%\\
DP-Adapters                                                 &    13.29                          &   13.07                          &           7.38                      &        80.84                     &        86.93                     &                                         90.15                                   &    84.20                         &              87.98               &        91.37                                                                    &    86.17                         &    90.28                         &    94.36                                                                        &   92.87                          &     95.33                        &   95.82                              &1.86\%\\

PromptDPSGD                                                    &   12.44                          &  12.12                           &         7.25                        &      81.16                       &      87.13                       &                                            90.48                                &     83.58                        &  88.46                           &                                                 91.22                           &  87.23                           &       90.87                      &     94.11                                                                       &    93.13                         &      95.29                       &  95.93                               &1.92\%\\ 
DP-MemArc$_\text{side}$ &     \textbf{6.18}                        &        \textbf{6.23}                     &                  \textbf{5.66}               &       81.30                      &  \textbf{87.16}                           &                             \textbf{90.91}                                               &               \textbf{84.56}              &     \textbf{88.92}                        &     91.66                                                                       &      86.95                       &     \textbf{91.56}                        &       \textbf{94.40}                                                                     &        \textbf{93.60}                     &      \textbf{95.44}                      &    \textbf{95.94}                             &2.10\%\\ \midrule
\multicolumn{17}{c}{\textbf{\textit{$\text{Differential Private on Fixed Parameter Transfer Learning}$}}} \\ \midrule
DP-Full FT                                                   &  26.12                           & 26.83                            &   10.93                         &  51.45                      &  84.23                      &\textbf{90.65}                                                                       &   61.37                     &    84.98                    &  \textbf{92.30}                                                                     &   59.55                     &   84.48                     &   \textbf{95.13}                                                                    &   75.74                     &  86.20                      &  \textbf{96.18}                          &100\%\\ 
DP-BiTFiT                                                    &   \textbf{5.12}                          &   5.88                          &             4.82                    &     75.36                        &        83.74                     &                                       89.19                                     &      78.92                       &  85.20                           &              90.65                                                              & 83.43                            &       87.57                      &      93.56                                                                      &      89.12                       &        93.02                     &         95.38                        &0.08\%\\

DP-MemArc$_\text{rev}$   &     5.48                        &     \textbf{5.65}                        &              \textbf{4.78}                   &     \textbf{80.29}                        &     \textbf{86.12}                        &                                                 90.21                           &              \textbf{82.57}              &    \textbf{88.12}                         &        91.25                                                                    &     \textbf{85.89}                        &    \textbf{90.31}                         &        94.10                                                                    &    \textbf{91.78}                         &     \textbf{93.89}                        &  95.32                               &3.92\%\\ \bottomrule
\end{tabular}
}
\caption{Experiments on the RoBERTa-large model. We evaluate the accuracy(\%) results and profile to compute the training memory(GB) with privacy constraints at \(\epsilon = 1.6, 8, \infty\). We propose two DP-MemArc architectures as novel efficient memory privacy fine-tuning schemes. Adaptive and Fixed are used to differentiate whether the trainable parameters can be adjusted.}
\label{tab:main:roberta}
\end{table*}

\section{Experimental Setup}

We designed a series of experiments covering different models and datasets to evaluate the performance of our methods. The specific experimental design is as follows.

\vspace{5pt}

{
\setlength{\parindent}{0cm}

\textbf{Models.}
We used the RoBERTa-large \cite{liu2019roberta}, GPT-2-large \cite{radford2019language} model as our base models. These models will be fine-tuned according to the corresponding downstream tasks, and the performance of the fine-tuned models will be evaluated under different privacy constraints.

}

\vspace{5pt}

{
\setlength{\parindent}{0cm}

\textbf{Baselines.} 
We compare the two methods against multiple baselines, including DP-LoRA \cite{hu2021lora, yu2021differentially}, DP-Adapter \cite{houlsby2019parameter, yu2021differentially}, DP-BiTFiT \cite{bu2024differentially, zaken2022bitfit}, and PromptDPSGD \cite{duan2024flocks, lester2021power}. These methods are all privacy-preserving fine-tuning approaches with opacus DP \cite{yousefpour2021opacus}, and we test them on the same training data to ensure fairness of comparison.
}

\vspace{5pt}

{
\setlength{\parindent}{0cm}

\textbf{Datasets.}
We conduct experiments on five datasets. Four from the GLUE benchmarks \cite{wang2018glue}, which cover different NLP tasks. \textbf{MNLI}: the MultiGenre Natural Language Inference Corpus. \textbf{QQP}: the Quora Question Pairs2 dataset. \textbf{QNLI}: the Stanford Question Answering dataset. \textbf{SST2}: the Stanford Sentiment Treebank dataset. We also select an NLG task \textbf{E2E} dataset \cite{duvsek2019evaluating}, which is to generates texts to evaluate a restaurant, to evaluate the quality of the model in generation tasks under privacy constraints. 
}

\vspace{5pt}

\begin{table*}[!t]
    \resizebox{\linewidth}{!}{    
\begin{tabular}{@{}c|ccccccccccccccc|c@{}}
\toprule
\multirow{2}{*}{}                                            & \multicolumn{3}{c}{Memory(GB)$\downarrow$}                                                              & \multicolumn{3}{c}{BLEU$\uparrow$}                                                                                                               & \multicolumn{3}{c}{Rouge-L$\uparrow$}                                                                                                                & \multicolumn{3}{c}{Perplexity$\downarrow$}                                                                                                                 & \multicolumn{1}{c}{\multirow{2}{*}{Trainable param(\%)}}                                                                \\ \cmidrule(lr){2-4} \cmidrule(lr){5-7} \cmidrule(lr){8-10} \cmidrule(lr){11-13}  
                                                             & $\epsilon = 1.6$ & $\epsilon = 8$ & $\epsilon = \infty$ & $\epsilon = 1.6$ & $\epsilon = 8$ & $\epsilon = \infty$ & $\epsilon = 1.6$ & $\epsilon = 8$ & $\epsilon = \infty$ & $\epsilon = 1.6$ & $\epsilon = 8$ & $\epsilon = \infty$ &  \\ \midrule
\multicolumn{14}{c}{\textbf{\textit{$\text{Differential Private on Adaptive Parameter Transfer Learning}$}}} \\ \midrule
DP-LoRA   &22.21&21.72&13.73&65.4&67.1&69.1&64.4&68.3&72.1&2.42&2.45&2.31&1.15\%\\
DP-Adapters  &23.68&24.12&14.55&65.2&66.9&\textbf{69.2}&\textbf{64.9}&68.2&71.9&2.44&2.35&2.28&1.16\%\\
PromptDPSGD        &22.12&20.96&14.18&64.2&66.5&69.1&65.0&68.3&72.0&2.60&2.54&2.39&1.33\%\\ 
DP-MemArc$_\text{side}$ &\textbf{11.68}&\textbf{11.44}&\textbf{10.17}&\textbf{66.4}&\textbf{68.2}&68.9&64.6&\textbf{68.5}&\textbf{72.7}&\textbf{2.32}&\textbf{2.38}&\textbf{2.24}&1.28\%\\ \midrule
\multicolumn{14}{c}{\textbf{\textit{$\text{Differential Private on Fixed Parameter Transfer Learning}$}}} \\ \midrule
DP-Full FT  &58.96&62.23&20.45&62.2&66.8&69.3&63.4&67.8&\textbf{72.6}&\textbf{2.46}&\textbf{2.23}&\textbf{1.85}&100\%\\
DP-BiTFiT           &9.59&\textbf{9.71}&8.62&61.7&65.2&68.6&62.9&66.4&71.3&2.83&2.58&2.77&0.05\%\\
DP-MemArc$_\text{rev}$  &\textbf{9.45}&9.88&\textbf{8.39}&\textbf{65.1}&\textbf{66.1}&\textbf{69.8}&\textbf{64.2}&\textbf{68.1}&71.6&2.71&2.65&2.58&2.15\%\\ \bottomrule
\end{tabular}
}
\caption{Experiments on the GPT-2-large model. We evaluate the BLEU(\%), Rouge-L(\%) and Perplexity scores results on E2E dataset and profile to compute the training memory(GB) with privacy constraints at \(\epsilon = 1.6, 8, \infty\). }
\label{tab:main:gpt2}
\end{table*}

{
\setlength{\parindent}{0cm}

\textbf{Implementation Details.}
To standardize the training process, we partition each dataset as follows: The text classification dataset includes 50k samples for training, 1k samples for validation, and the remaining data for testing. The E2E dataset includes 42061 samples for training and 4672 samples for validation.	We set different privacy constraint conditions specifically as \(\epsilon = \{1.6, 8, \infty\}\) and $\delta = 1/|\mathcal{D}_{train}|$ to assess performance variations among different methods under these constraints. We chose a learning rate of \texttt{5e-4} and used DP-Adam optimizer as the default optimizer for the model, while DP-SGD optimizer is employed for PromptDPSGD. For evaluation metrics, we utilize a profiler to track the model's training memory usage, evaluating the mean memory consumption during training. Default LoRA and Adapters ranks are set to $r = 64$. For text classification tasks, we compare accuracy. For generation tasks, we employed perplexity, BLEU \cite{papineni2002bleu}, and ROUGE-L \cite{lin2004rouge} as evaluation metrics to comprehensively assess generation quality. In our experiments, we conduct training with a batch size of 32 and sequence length of 128 in FP16. 

}

\section{Experiments}

\subsection{Main Results}

We evaluate various baseline methods on multiple task datasets and organized the results of RoBERTa and GPT2 separately according to the task type.

\vspace{5pt}

{
\setlength{\parindent}{0cm}

\textbf{Text classification on RoBERTa-large.} As shown in Table~\ref{tab:main:roberta}, the two DP-MemArc methods demonstrate competitive performance on text classification tasks using the RoBERTa-large model. 

\vspace{3pt}

(1) The side network design achieves the best results compared to other adaptive baseline methods in most of the accuracy evaluations. The average performance on DP-MemArc$_\text{side}$ is similar to DP-LoRA, but the side network design method requires less training memory than DP-LoRA. 

\vspace{3pt}

(2) Specifically, compared to the performance of DP-LoRA under privacy constraints, our DP-MemArc$_\text{side}$ achieves nearly $2 \sim 3 \times$ optimization in training memory. Simultaneously, we can observe that when further memory savings during training are required, the reversible network design of DP-MemArc offers an ideal choice. 

\vspace{3pt}

(3) Compared to the current most memory-efficient method, DP-BiTFiT, our method consistently performs better in downstream tasks while maintaining similar training memory usage. This indicates that DP-MemArc$_\text{rev}$ can better learn the characteristics of downstream tasks and perform gradient clipping based on computable activation function values while preserving privacy. 

\vspace{3pt}

(4) In terms of average performance, DP-MemArc$_\text{rev}$ improves accuracy by an average of \textbf{+3.1\%} compared to DP-BiTFiT and performs better in scenarios with higher privacy constraints $\epsilon$, suggesting that the model better captures the gradient changes of the training data and adapts to downstream tasks.

}

\vspace{6pt}

{
\setlength{\parindent}{0cm}

\textbf{Text Generation on GPT-2-large.} 
For generative tasks, we employ three metrics to assess the quality of animal generation and simultaneously utilize profiles to record the changes in training memory. Experiments on Table~\ref{tab:main:gpt2} indicate that our approach demonstrates performance comparable to text classification tasks in generative tasks. 

\vspace{3pt}

(1) Our side network design excels in perplexity performance compared to other differential privacy parameter tuning methods. Additionally, DP-MemArc$_\text{side}$ shows outstanding performance on the BLEU metric. Comparing our method under differential privacy, when the parameter $\epsilon$ is set to 1.6 indicating higher privacy demands, performance in the BLEU metric only drops by 3.5\%. This suggests our method better learns the characteristics and paradigms of generative tasks, yielding relatively accurate outputs.

\vspace{3pt}

(2) Compared to DP-BiTFiT, reversible network design exhibits competitive training memory consumption requirements, with DP-MemArc$_\text{rev}$ maintaining strong performance. This approach maintains relatively stable task accuracy under highly constrained training memory conditions.

\vspace{3pt}

(3) Compared to full differential privacy fine-tuning, DP-MemArc$_\text{rev}$ saves approximately $6 \sim 8 \times$ the training memory in high privacy $\epsilon = 1.6$ scenarios. These results underscore the promising outlook of our proposed DP-MemArc framework for generative tasks, maintaining lower training memory requirements even at larger batch sizes.

}

\subsection{Analysis}

We conduct a deep analysis of two DP-MemArc methods and perform ablation experiments on the corresponding modules, including the differential private algorithm and alternative model setting.

\vspace{5pt}

{
\setlength{\parindent}{0cm}

\textbf{Book-Keeping in DP-MemArc.} In the setup of these two architectures, we use the BK-MixOpt method for differential privacy training. BK-MixOpt reduces the required training memory by using Ghostnorm to compute the normalized formulation. To evaluate the impact of different differential privacy methods during the training process, we conducted experiments on these two model designs and measured the average memory consumption during the training process. The results are shown in Table~\ref{tab:bk}.

}

BK-MixOpt exhibits the best performance in the following scenarios. From the ablation experiments, the BK-MixOpt method reduces training memory consumption by $1.5 \sim 2 \times$ in privacy-preserving computation. This highlights the importance of using BK-MixOpt within our framework. When there are no privacy constraints as $\epsilon = \infty$, all three methods degrade into the standard gradient descent process. Under the condition of privacy constraints, if the Opcaus method of calculating gradients for each sample is adopted, the time complexity for calculating the sample gradient in a single layer under the two architectures DP-MemArc$_\text{side}$ and DP-MemArc$_\text{rev}$ is $O(Bpd/64)$ and $O(4Bpr)$. This still requires a considerable amount of computation time, and in DP-MemArc$_\text{side}$, the gradient calculation for the upsampling and downsampling matrices also needs to be considered.


\begin{table}[]
    \resizebox{\linewidth}{!}{    
\begin{tabular}{@{}c|lcc@{}}
\toprule
                              & Privacy Constrains & DP-MemArc$_\text{side}$ & DP-MemArc$_\text{rev}$ \\ \midrule
\multirow{3}{*}{Opcaus}       & $\epsilon = 1.6, \delta = 2 \times 10^{-5}$   &   7.45                   &        10.66              \\
                              & $\epsilon = 8.0, \delta = 2 \times 10^{-5}$   &    7.33                  &      10.98                \\
                              & $\epsilon = \infty, \delta = 2 \times 10^{-5}$   &           5.60           &       4.82               \\ \midrule
\multirow{3}{*}{GhostClip}    & $\epsilon = 1.6, \delta = 2 \times 10^{-5}$   &    9.72                  &      8.52                \\
                              & $\epsilon = 8.0, \delta = 2 \times 10^{-5}$   &              9.54        &       8.43               \\
                              & $\epsilon = \infty, \delta = 2 \times 10^{-5}$   &           5.72           &      4.75                \\ \midrule
\multirow{3}{*}{BK-MixOpt} & $\epsilon = 1.6, \delta = 2 \times 10^{-5}$   &   6.18                   &       5.48               \\
                              & $\epsilon = 8.0, \delta = 2 \times 10^{-5}$  &   6.23                   &    5.65                  \\
                              & $\epsilon = \infty, \delta = 2 \times 10^{-5}$   & 5.66                     &    4.78                  \\ \bottomrule
\end{tabular}
}
\caption{Evaluations of Different DP methods on DP-MemArc. }
\label{tab:bk}
\end{table}

\vspace{5pt}

{
\setlength{\parindent}{0cm}

\textbf{Reversible Network Functions.} 
In the design of DP-MemArc$_\text{rev}$, we include two sub-functions that are used to achieve the reversible design of reversible networks. Section~\ref{sec:reverse} elaborates on the principles of the reversible network's inversion. Therefore, we can modify the internal design while ensuring that each sub-function fulfills its respective role. To further understand the differences between various designs, we fix the sub-function $\mathcal{G}$ and change the internal architecture of sub-function $\mathcal{F}$, replacing it with different parameter-efficient fine-tuning(PEFT) methods. In the privacy scenario, we select different methods and incorporate them with DP-MemArc$_\text{rev}$ in terms of accuracy and training memory consumption.

}

\begin{figure}[tp]
    \centering
    \includegraphics[width=1\linewidth]{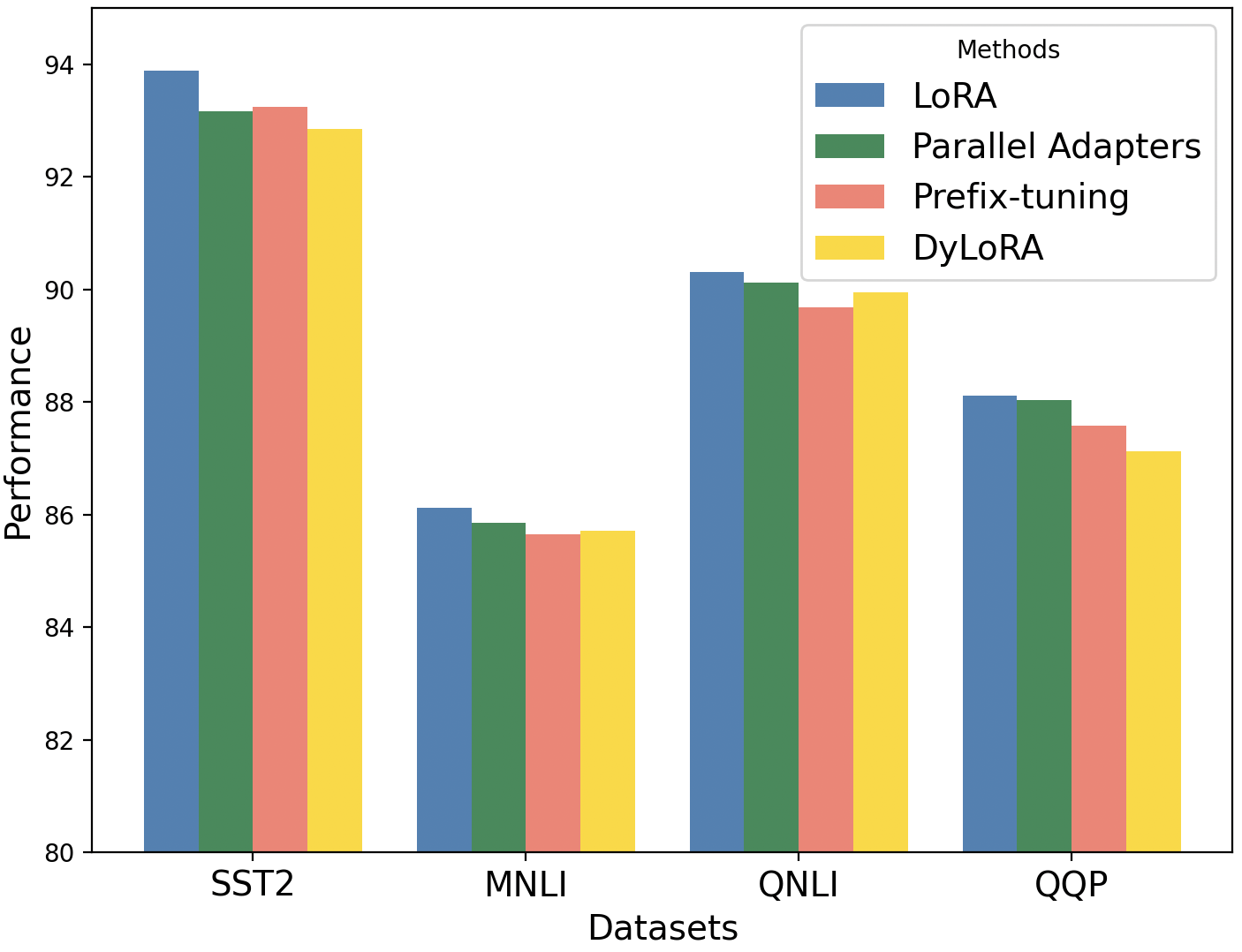}
    \caption{Performance of different reversible network sub-function $\mathcal{F}$ design. The private constraint is $\epsilon = 8.0$.}
    \vspace{-0.15cm}
    \label{fig:sc}
\end{figure}

We have selected several classic and efficient parameter fine-tuning methods to replace the subfunction $\mathcal{F}$ here, including LoRA \cite{hu2021lora}, Parallel Adapters \cite{he2021towards}, Prefix tuning \cite{li2021prefix} and dyLoRA \cite{valipour2023dylora}, and set the constraint $\epsilon = 8.0$. The result is shown in Figure~\ref{fig:sc}.

LoRA is superior to other candidate architectures as a reversible network sub-function. Compared to other methods, using $\mathcal{F} = \F(\W_\text{p} + LR, \theta + \Delta\theta; x_i^2)$ results in a slight \textbf{+0.71} improvement in accuracy. Given the simplicity of LoRA's network architecture and the similarity in training memory usage across various methods, we finally adopt LoRA as the reversible network sub-function for DP-MemArc$_\text{rev}$.

\subsection{Time Efficiency}

To better validate the time efficiency theory in Table~\ref{tab:block complexity}, we compare the time efficiency of our method with the baseline method to ensure that DP-MemArc maintains consistent time consumption while being memory-efficient. Since the inference speed of a model is positively correlated with the size of the training parameters in most cases, when we set the same number of trainable parameters, the impact of parameter size is reduced, and the difference in consumption is mainly reflected in the different frameworks. The corresponding results are shown in the Table~\ref{tab:time}. The experimental results indicate that compared to other baseline methods, DP-MemArc is more efficient in reasoning across various downstream tasks.

\subsection{Training Scale Analysis}

To understand and compare the training process and accuracy variations of different methods under differential privacy, we use checkpoints to record the training process of the model. We test three methods: DP-LoRA, DP-MemArc$_\text{side}$, and DP-MemArc$_\text{rev}$ on the GPT2-large model, evaluating their BLEU scores. From the results, DP-MemArc$_\text{side}$ and DP-MemArc$_\text{rev}$ require more training steps to reach stable values compared to DP-LoRA. Considering the architecture of the models themselves, DP-MemArc$_\text{side}$ needs to be tuned for the entire side network to adapt to the corresponding time for downstream tasks. Training the low-rank matrices of DP-LoRA is relatively simpler. As for the reversible network, due to the use of approximation methods for learning, more training data helps to mitigate the performance loss caused by approximation by adjusting the reversible gradients.

\begin{table}[]
    \resizebox{\linewidth}{!}{  
\begin{tabular}{@{}c|ccc@{}}
\toprule
                 & Time Complexity(sec) & Memory Complexity(GB) & QQP scores↑ \\ \midrule
DP-Full FT       & 626.51               & 26.83                 & 84.98       \\
DP-LoRA          & 295.62               & 12.68                 & 88.73       \\
DP-Adapters      & 301.96               & 13.07                 & 87.98       \\
DP-BiTFiT        & 285.98               & 5.88                  & 85.20       \\
PromptDPSGD      & 293.65               & 12.06                 & 88.31       \\ \midrule
DP-MemArc$_\text{side}$ & 291.56               & 6.23                  & \textbf{88.92}       \\
DP-MemArc$_\text{rev}$  & \textbf{284.69}              & \textbf{5.65}                 & 88.12       \\ \bottomrule
\end{tabular}
}
\caption{Experiments on time efficiency for different methods on Roberta-large. The privacy constraints are set to $\epsilon = 8$.}
\label{tab:time}
\end{table}

\section{Related Work}

\subsection{Differential Private Fine-tuning}

To ensure the privacy needs of the model, differential privacy fine-tuning methods offer a feasible solution with strong theoretical guarantees \cite{abadi2016deep, song2013stochastic}. In terms of model structure, PEFT methods can be transferred to differential privacy schemes \cite{yu2021differentially, bu2024differentially, xu2024dp}. In methods design, the selected differential privacy \cite{shi2022selective, shi2022just} approach can provide stronger differential privacy constraints more specifically for designated information. In algorithm design, it includes a series of studies \cite{rochette2020efficient, du2023dp} on the computational graph during the differential privacy propagation process. Techniques like Ghostnorm \cite{goodfellow2015efficient, li2021large} and Book-Keeping \cite{bu2023differentially} provide unified batch norm computation and batch processing for gradient clipping. Although differential privacy offers very strong theoretical protections, reducing the memory requirements for training under differential privacy scenarios remains a significant challenge \cite{du2023dp}. 

\subsection{Parameter Efficient Transfer Learning}

To reduce the demand for computational resources during training, parameter-efficient fine-tuning methods are applied to transfer learning. Common methods include training low-rank matrices \cite{hu2021lora, valipour2023dylora, dettmers2024qlora}, adding adapters \cite{houlsby2019parameter, he2021towards}, and performing prefix tuning \cite{li2021prefix, liu2022p} or prompt tuning \cite{lester2021power} on the inputs of the original model. While most parameter-efficient fine-tuning methods reduce time and space consumption, they still require significant training memory due to the state of activation functions \cite{sung2022lst, liao2023make}. Our framework offers side networks and reversible networks designs to reduce the memory required during training.

\begin{figure}[tp]
    \includegraphics[width=1\linewidth]{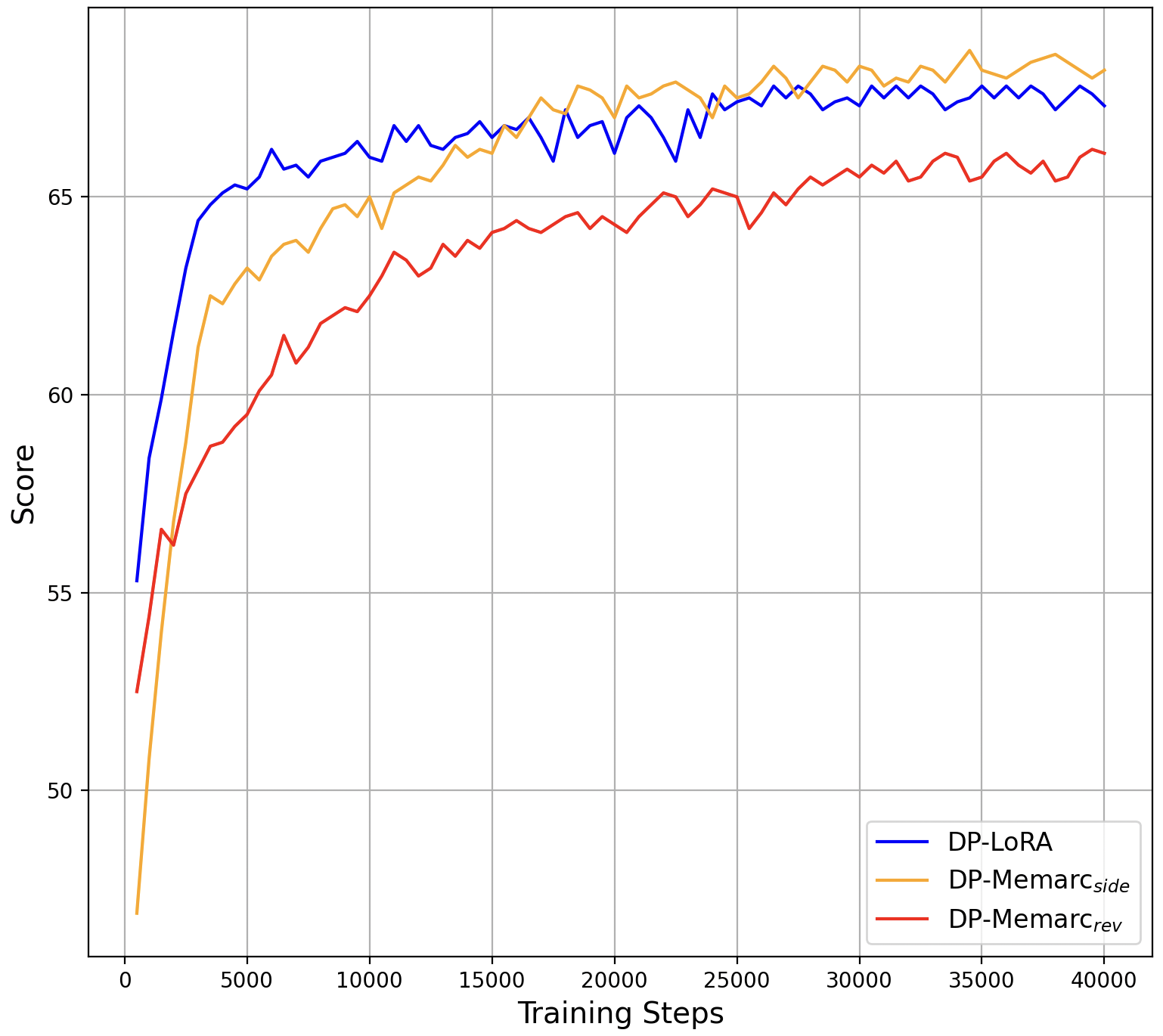}
    \caption{The experiment of training steps is conducted on the E2E dataset. }
    \vspace{-0.3cm}
    \label{fig:training_acc}
\end{figure}

\section{Conclusion}

In this paper, we introduce a framework called DP-MemArc, which encompasses two methods aimed at addressing the issue of excessive memory consumption during training in privacy-sensitive scenarios. In this process, we reduce the training memory consumption of models in privacy environments using the BK method. With DP-MemArc, LLMs can perform downstream tasks under corresponding privacy constraints across various tasks. We hope that our method will contribute to future private efficient memory optimization for fine-tuning LLMs.

\section{Acknowledgements}

This work was partly supported by the NSFC under No. 62402418 and No. 62102360. This work was also partly supported by the Key R\&D Program of Ningbo under No. 2024Z115.

\bibliography{aaai25}

\end{document}